\documentclass[aps,pre,showpacs,twocolumn,amsmath,amssymb]{revtex4-1}
\usepackage{graphicx}% Include figure files
\usepackage{dcolumn}% Align table columns on decimal point
\usepackage{bm}% bold math
\usepackage{amsmath}
\usepackage{amssymb}
\usepackage{latexsym}
\usepackage{epsfig}
\usepackage{amsbsy}
\usepackage{array}
\usepackage{amssymb}
\usepackage{setspace}
\usepackage{bm}
\begin{document}
%%%%%%%%%%%

\title{Extracting work from the magnetic field coupled Brownian particles}

\author{Tian Chen$^{1,2,3}$\footnote{Email:chentian10@mails.tsinghua.edu.cn}}

\author{Xiang-Bin Wang$^{1,3,4}$\footnote{Email:xbwang@mail.tsinghua.edu.cn}}

\author{Ting Yu$^2$\footnote{Email:ting.yu@stevens.edu}}

\affiliation{$^{1}$State Key Laboratory of Low
Dimensional Quantum Physics, Department of Physics, Tsinghua University, Beijing 100084,
People's Republic of China\\
$^2$Center for Controlled Quantum Systems and the Department of Physics
and Engineering Physics, Stevens Institute of Technology, Hoboken,
New Jersey 07030, USA\\
$^3$Synergetic Innovation Center of Quantum Information and Quantum Physics, University of Science and Technology of China, Hefei, Anhui 230026, People's Republic of China\\
$^4$Jinan Institute of Quantum Technology, Shandong
Academy of Information and Communication Technology, Jinan 250101,
People's Republic of China
}
\begin{abstract}
Thermodynamics of the magnetic field coupled Brownian particles is studied. We show that in the presence of the magnetic field,
work can be extracted from the reservoir even when the measurement operation and the potential change operation are applied in different spatial
directions.  In particular, we show that more work can be extracted if the measurements are applied in two different directions simultaneously. In all these cases, we
show that the generalized second law involving the measurement information and potential change is satisfied. In addition, we show
how  the continuous potential change and measurement position affect the work extraction.
%Keywords: coupled Brownian particles, work
\end{abstract}


%\begin{keywords}
% Keywords: Transition-metal Aluminides; Slater Pauling rule;
% Half-metallicity; first-principles
\maketitle
\section{Introduction}
Extracting work from a system (reservoir) has attracted a widespread research interest recently \cite{Seifert_RPP, Nori_RMP}. The original Jarzynski equality \cite{Jarzynski97, Jarzynski99PRE, Esposito09RMP, Campisi11RMP, Mukamel06PRE, Jimenez10JPA, Hu_PRE12, Leggio13PRE} culminating in the recent developments states that, for a finite time nonequilibrium process, $\langle e^{-\beta W}\rangle=e^{-\beta\Delta F}$, this means that $W\geqslant\Delta F$, where $W$ is the work exerted on the system, and $\Delta F$ is the equilibrium free energy difference between the initial and final states. More recently, an information-theoretic concept -- relative entropy has been introduced \cite{Ueda_PRL08, Marin_PRE08, Jacobs_PRA09}, and the original Jarzynski equality is then extended to a new context where the measurement process and the information contribution from the measurement have been incorporated to produce a generalized Jarzynski equality $\langle e^{-\beta(W-\Delta F)-I}\rangle=1$, where the quantity $I$ is the obtained information measured by the relative entropy. Then the generalized second law is written as \cite{Ueda_PRL10, Toyabe_Nat10, Granger11PRE, D. Abreu11, Ueda_PRL12, Esposito_PRL13, SagawaNJP13, MunakataJSM2013, Funo13},
\begin{equation}
W\geq \Delta F-I.\label{one}
\end{equation}

From Eq.~(\ref{one}), it becomes clear that, when the parameters are chosen appropriately, the thermal equilibrium states of the system remain the same at the initial and final times, hence the applied work can be negative due to the presence of the  information contribution. Physically, it implies that one can extract work from the reservoir.  Recently several models have been studied including Szilard machine \cite{Szilard29, Kim_PRL} where the work can be extracted from the thermal reservoir by employing the information of the particle's position in the box,  and the Brownian particle model \cite{Ueda_PRL10, D. Abreu11, M. Bauer12} where one can show that the Brownian particle can be driven
by using the information of the measurement outcomes of the particle's position \cite{Ueda_PRL10}. Furthermore,  a more sophisticated Brownian motion model with variable potential's stiffness and potential center together with a measurement process has been studied to show how the maximum work can be extracted from the reservoir  \cite{D. Abreu11, M. Bauer12}. Other interesting developments that are devoted to energy and information extraction include the transporting particles devices through writing the information on the bit tape  \cite{D. Mandal, Barato13EPL, HorowitzPRL13}, the electron transport \cite{Esposito_PRL13} where one electronic state of a quantum dot is used as a detector, which only affects the entropy production of the probed conventional single electron transistor (SET) etc. \cite{Horowitz_NJP, Horowitz_EPL, Esposito_EPL11, Horowitz13arXiv}.

The purpose of this paper is to study the work extracting from the system consisting of the magnetic field coupled Brownian particles by incorporating  the obtained
information from measurement processes based on a model for the dynamics of the plasma diffusion \cite{Jimenez10JPA, J. B. Taylor1962, Bastida2012, R. M. Velasco2013}. More specifically,
we will consider the work and the measurement relation and explore the work extraction in four interesting setups. First, we consider the measurement operation and the potential change operation in the same direction. Then,  we consider a different setup where the two operations are executed in different directions. Next,  we consider the work extraction when two independent measurements are considered. Finally, we discuss the extracted work when the potential change operations are made in both two directions, and the measurement takes place only on one direction. We find that, for all four setups, when the measurement effects are taken into account,  the generalized second law is valid even when the measurement operation and the changing potential operation are spatially separated. It is interesting to note that more work can be extracted when two independent measurements are performed. In addition, we have studied the magnetic field effect on the thermodynamics of this coupled particles system.

The organization of our work is as follows,  Sec.~\ref{II} introduces the coupled Brownian particles model, the initial system state, and the definitions of
the related thermodynamic quantities. In Sec.~\ref{III} we study the effect of the outcomes of different position measurements, measurement accuracies,
and frequencies of the potential. We conclude in Sec.~\ref{IV}.

\section{model}\label{II}
%%%%%%%%%%%%

\textit{Equation of motion}. The dynamics of the magnetic field coupled Brownian particles can be described by a stochastic Langevin equation \cite{Jimenez10JPA, J. B. Taylor1962, Bastida2012, R. M. Velasco2013},
\begin{equation}
M\frac{d\vec{u}}{dt}=-\gamma\vec{u}+\frac{q}{c}\vec{u}\times\vec{B}-k(\vec{x}-\vec{x}_{0})+\vec{\mu}(t),
\end{equation}
with $\vec{u}=(v_{x},v_{y})$, $\vec{x}=(x,y)$, $\vec{x}_{0}=(x_{0},y_{0})$, the noise term $\vec{\mu}(t)=(\mu_{x}(t),\mu_{y}(t))$  satisfying $\langle\mu_{i}(t)\mu_{j}(t')\rangle=\gamma k_{B}T\delta_{ij}\delta(t-t')$ $(i=x,y)$ where $k_{B}$ is the Boltzmann's constant and $T$ is the thermal bath temperature.  $\gamma$ is the friction coefficient. The potential fields along $x$ and $y$ directions are $V_{x}=\frac{1}{2}k_{x}(x-x_{0})^{2}$, $V_{y}=\frac{1}{2}k_{y}(y-y_{0})^{2}$, respectively.  The magnetic field $\vec{B}$ is applied along the $z$ direction.

The above equation
can be explicitly written into a set of coupled differential equations along the $x$, $y$ directions, respectively,
%\begin{equation}
%\begin{split}
%\left\{\begin{array}{l}
%\dot{v}_{x}=-\frac{\gamma}{M}v_{x}-\frac{k_{x}}{M}(x-x_{0})+\frac{q}{cM}v_{y}B+\frac{\mu_x(t)}{M},\\
%\dot{x}=v_{x},\\
%\dot{v}_{y}=-\frac{\gamma}{M}v_{y}-\frac{k_{y}}{M}(y-y_{0})-\frac{q}{cM}v_{x}B+\frac{\mu_y(t)}{M},\\
%\dot{y}=v_{y}.\\
%\end{array}\right.
%\end{split}
%\end{equation}
\begin{subequations}
\begin{align}
\dot{v}_{x}&=-\frac{\gamma}{M}v_{x}-\frac{k_{x}}{M}(x-x_{0})+\frac{q}{cM}v_{y}B+\frac{\mu_x(t)}{M},\\
\dot{x}&=v_{x},\\
\dot{v}_{y}&=-\frac{\gamma}{M}v_{y}-\frac{k_{y}}{M}(y-y_{0})-\frac{q}{cM}v_{x}B+\frac{\mu_y(t)}{M},\\
\dot{y}&=v_{y}.
\end{align}
\end{subequations}
Here, for simplicity, we set $M=1$, and employ the Fokker-Planck (FP) equation to describe the system dynamics \cite{Risken},
\begin{equation}\small
\begin{split}
\frac{\partial \mathcal{P}(x,v_{x},y,v_{y},t)}{\partial t}=&\{-\frac{\partial}{\partial x}v_{x}+\frac{\partial}{\partial v_{x}}(\gamma v_{x}+k_{x}(x-x_{0})-\frac{q}{c}v_{y}B)\\&+\gamma k_{B}T\frac{\partial^{2}}{\partial v_{x}^{2}}-\frac{\partial}{\partial y}v_{y}\\&+\frac{\partial}{\partial v_{y}}(\gamma v_{y}+k_{y}(y-y_{0})+\frac{q}{c}v_{x}B)\\&+\gamma k_{B}T\frac{\partial^{2}}{\partial v_{y}^{2}}\}\mathcal{P}(x,v_{x},y,v_{y},t).
\end{split}
\end{equation}
where $\mathcal{P}(x,v_{x},y,v_{y},t)$ denotes the probability distribution function of the system at time $t$.

In the overdamped approximation, the system dynamics can be simplified as,
%\begin{equation}
%\begin{split}
%\left\{\begin{array}{l}
%\gamma v_{x}=-k_{x}(x-x_{0})+\frac{q}{c}v_{y}B+\mu_{x}(t),\\
%\gamma v_{y}=-k_{y}(y-y_{0})-\frac{q}{c}v_{x}B+\mu_{y}(t),\label{oqbm01}
%\end{array}\right.
%\end{split}
%\end{equation}
\begin{subequations}
\begin{align}
\gamma v_{x}=-k_{x}(x-x_{0})+\frac{q}{c}v_{y}B+\mu_{x}(t),\\
\gamma v_{y}=-k_{y}(y-y_{0})-\frac{q}{c}v_{x}B+\mu_{y}(t).\label{oqbm01}
\end{align}
\end{subequations}
The FP equation can be written in a more transparent form if we use the following new notations:  $R=x-\frac{qB}{c\gamma}y$, and $S=\frac{qB}{c\gamma}x+y$, $H(t)=\frac{k_{x}(t)}{\gamma(1+(qB/c\gamma)^{2})}$, $J(t)=\frac{k_{y}(t)}{\gamma(1+(qB/c\gamma)^{2})}$, $N=\frac{k_{B}T}{\gamma}$, and $L=\frac{qB}{c\gamma}$, then the FP equation of the system distribution $\mathcal{P}(R,S,t)$ is given by,
\begin{equation}
\begin{split}
&\frac{\partial \mathcal{P}(R,S,t)}{\partial t}=\{H(t)\frac{\partial}{\partial R}(R+L\cdot S)+J(t)\frac{\partial}{\partial S}(S-L\cdot R)\\&-\frac{k_{x}(t)}{\gamma}x_{0}\frac{\partial}{\partial R}-\frac{k_{y}(t)}{\gamma}y_{0}\frac{\partial}{\partial S}+N\frac{\partial^{2}}{\partial R^{2}}+N\frac{\partial^{2}}{\partial S^{2}}\}\mathcal{P}(R,S,t).
\end{split}
\end{equation}

\textit{Analytical Solution}. The above FP equation can be solved when the Fourier transform is applied:  $\mathcal{P}(R,S,t)=\frac{1}{(2\pi)^{2}}\int dp\int dq e^{i(Rp+Sq)}\tilde{\mathcal{P}}(p,q,t)$. Note that the parameters transformation $\frac{\partial}{\partial R}\rightarrow ip$, $\frac{\partial}{\partial S}\rightarrow iq$, $R\rightarrow i\frac{\partial}{\partial p}$, $S\rightarrow i\frac{\partial}{\partial q}$.  We can decompose the probability function in the frequency domain as, $\tilde{\mathcal{P}}=\tilde{\mathcal{P}}_{1}\cdot\tilde{\mathcal{P}}_{2}\cdot\tilde{\mathcal{P}}_{3}$, with $\tilde{\mathcal{P}}_{1}=\exp(-ipf(t)-p^{2}g(t))$, $\tilde{\mathcal{P}}_{2}=\exp(-iqh(t)-q^{2}e(t))$, and $\tilde{\mathcal{P}}_{3}=\exp(-pq\kappa(t))$. Then it is easy to show that the  coupled equations for the coefficients $f(t)$, $h(t)$, $g(t)$, $e(t)$, and $\kappa(t)$ take the following forms,
%\begin{equation}\small
%\begin{split}
%\left\{\begin{array}{l}
%-i\dot{f}(t)=iH(t)f(t)+iH(t)L\cdot h(t)-i\frac{k_{x}(t)}{\gamma}x_{0},\\
%-i\dot{h}(t)=iJ(t)h(t)-iJ(t)L\cdot f(t)-i\frac{k_{y}(t)}{\gamma}y_{0},\\
%-\dot{g}(t)=2H(t)g(t)+H(t)L\cdot\kappa(t)-N,\\
%-\dot{e}(t)=2J(t)e(t)-J(t)L\cdot\kappa(t)-N,\\
%-\dot{\kappa}(t)=H(t)\kappa(t)+J(t)\kappa(t)+2H(t)L\cdot e(t)-2J(t)L\cdot g(t).\\
%\end{array}\right.
%\end{split}
%\end{equation}
\begin{subequations}\small
\begin{align}
-\dot{f}(t)&=H(t)f(t)+H(t)L\cdot h(t)-\frac{k_{x}(t)}{\gamma}x_{0},\\
-\dot{h}(t)&=J(t)h(t)-J(t)L\cdot f(t)-\frac{k_{y}(t)}{\gamma}y_{0},\\
-\dot{g}(t)&=2H(t)g(t)+H(t)L\cdot\kappa(t)-N,\\
-\dot{e}(t)&=2J(t)e(t)-J(t)L\cdot\kappa(t)-N,\\
-\dot{\kappa}(t)&=H(t)\kappa(t)+J(t)\kappa(t)+2H(t)L\cdot e(t)-2J(t)L\cdot g(t).
\end{align}
\end{subequations}
%and the probability $\mathcal{P}(R,S,t)$ can be obtained by the inverse Fourier transform of $\tilde{\mathcal{P}}(p,q,t)$. The quantity $\gamma$ is set $1$. When we have the form $H(t)$, $M$, $N$ and $L$, we can numerically get the probability $\mathcal{P}(x,y,t)$.
The inverse Fourier transform is applied, and we obtain the system distribution $\mathcal{P}(x,y,t)$ as,
\begin{equation}\small
\begin{split}
&\mathcal{P}(x,y,t)=\frac{1}{4\pi}\frac{1}{\sqrt{g(t)e(t)-(\frac{\kappa(t)}{2})^2}}\exp\{-\frac{1}{|4g(t)e(t)-\kappa(t)^2|}\\&\cdot(Q_{1}(x-A_{1})^{2}+Q_{2}(y-A_{2})^2+Q_3
(x-A_1)(y-A_2))\},\\
\end{split}
\end{equation}
with $A_1=\frac{f(t)+h(t)L}{1+L^2}$, $A_2=\frac{h(t)-f(t)L}{1+L^2}$, $Q_1=e(t)-\kappa(t)L+g(t)L^2$, $Q_2=g(t)+\kappa(t)L+e(t)L^2$, and $Q_3=\kappa(t)(L^2-1)-2L(e(t)-g(t))$.

\textit{Initial state}.  The initial state distributions along the $x$ and $y$ directions are given below, respectively,
%\begin{equation}
%\begin{split}
%\left\{\begin{array}{l}
%\mathcal{X}(x,0)=\frac{1}{\sqrt{2\pi u^{2}_{x}(0)}}\exp(-\frac{(x-d_{x}(0))^{2}}{2u^{2}_{x}(0)}),\\
%\mathcal{Y}(y,0)=\frac{1}{\sqrt{2\pi u^{2}_{y}(0)}}\exp(-\frac{(y-d_{y}(0))^{2}}{2u^{2}_{y}(0)}),\\
%\end{array}\right.
%\end{split}
%\end{equation}
\begin{subequations}
\begin{align}
%\left\{\begin{array}{l}
\mathcal{X}(x,0)=\frac{1}{\sqrt{2\pi u^{2}_{x}(0)}}\exp(-\frac{(x-d_{x}(0))^{2}}{2u^{2}_{x}(0)}),\\
\mathcal{Y}(y,0)=\frac{1}{\sqrt{2\pi u^{2}_{y}(0)}}\exp(-\frac{(y-d_{y}(0))^{2}}{2u^{2}_{y}(0)}),
%\end{array}\right.
\end{align}
\end{subequations}
where  $d_{x,0}(d_{y,0})$ and  $u^{2}_{x,0}(u^{2}_{y,0})$ denote the initial average and variance in $x(y)$ prior to a measurement, respectively.

\textit{Work}. If the system Hamiltonian is parametrized by a single quantity $\lambda$, the work performed during the time interval $[0,$ $t]$ along one trajectory $z_{t}$ is given by~\cite{Jarzynski97},
\begin{equation}
w(z_{t})=\int_{0}^{t}d\tau\dot{\lambda}\frac{\partial\mathcal{H}(\lambda)}{\partial\lambda}(z_{\tau}).
\end{equation}
Here, $\mathcal{H}(\lambda)$ is the system Hamiltonian. Then averaging over all the possible trajectories with the probability $\mathcal{P}(x,y,t)$,
one gets the mean work $W$ \cite{Jarzynski97, D. Abreu11}. For the Brownian particle model considered in this paper,
the time-dependent terms are $k_{x}(t)$ and $k_y(t)$, and the total work during the time interval $[0,$ $t]$ can be calculated as,
\begin{equation}\small
\begin{split}
W=&\frac{1}{2(1+L^2)^4}\int_{0}^{t}d\tau\{\dot{k_{x}}(\tau)[f^{2}(\tau)+2g(\tau)+L^{2}(h^{2}(\tau)+2e(\tau))\\
&+2L(\kappa(\tau)+f(\tau)h(\tau))-2x_{0}(1+L^{2})(f(\tau)+Lh(\tau))\\&+x_{0}^{2}(1+L^{2})^{2}]+\dot{k_y}(\tau)[h^2(\tau)+2e(\tau)
+L^2(f^2(\tau)+2g(\tau))\\&-2L(\kappa(\tau)+f(\tau)h(\tau))-2y_0(1+L^2)(h(\tau)-f(\tau)L)\\&+y_0^2(1+L^2)^2]\}.
\end{split}
\end{equation}

%For our Brownian particle model discussed below, only the $x$ direction potential is time dependent($k_{x}(t)$), and the work value during the time interval $t\in[0,$ $t]$ can be calculated as \cite{Jarzynski97, D. Abreu11},
%\begin{equation}\small
%\begin{split}
%W=&\frac{1}{2(1+L^2)^4}\int_{0}^{t}d\tau\dot{k_{x}}(\tau)[f^{2}(\tau)+2g(\tau)\\
%&+L^{2}(h^{2}(\tau)+2e(\tau))+2L(\kappa(\tau)+f(\tau)h(\tau))\\
%&-2x_{0}(1+L^{2})(f(\tau)+Lh(\tau))+x_{0}^{2}(1+L^{2})^{2}].
%\end{split}
%\end{equation}

\section{Work extraction with measurement and potential change}
\label{III}
%%%%%%%%%%%%%%%%%%%%%%%%%%%
We will consider four different setups where the work extraction and the generalized second law will be investigated. The first setup involves a measurement operation and the potential change operation in the different directions while the second one concerns the special case where the measurement operation and the potential change operation
take place in the same direction. In the third setup, two measurements in different directions are considered. For these three setups, only one direction potential is time-dependent ($k_{x}(t)$ or $k_y(t)$). The fourth setup contains the potential change operations in both directions $x$ and $y$, and the measurement operation is executed along only one direction.

Specifically, for the first setup, we assume that  the measurement is along the $y$ direction and the outcomes satisfy the Gaussian distribution
$\mathcal{Y}(y_{m}|y)=\frac{1}{\sqrt{2\pi u_{y,m}^{2}}}\exp(-\frac{(y_{m}-y)^{2}}{2u_{y,m}^{2}})$ where the coefficients $y_{m}$ and $u_{y,m}$ stand for
the position measurement outcome and accuracy, respectively. The more accurate the measurement, the smaller the $u_{y,m}$ values.
The $y$ direction distribution after the measurement is then given by,
\begin{equation}\small
\mathcal{Y}(y|y_{m})=\frac{1}{\sqrt{2\pi\frac{u_{y}^{2}(0)u_{y,m}^{2}}{u_{y}^{2}(0)+u_{y,m}^{2}}}}\exp(-\frac{(y-\frac{y_{m}u_{y}^{2}(0)+d_{y}(0)
u_{y,m}^{2}}{u_{y}^{2}(0)+u_{y,m}^{2}})^{2}}{2\frac{u_{y}^{2}(0)u_{y,m}^{2}}{u_{y}^{2}(0)+u_{y,m}^{2}}}),
\end{equation}
Note that the distribution along the $x$ direction is $\mathcal{X}(x,0)=\frac{1}{\sqrt{2\pi u^{2}_{x}(0)}}\exp(-\frac{(x-d_{x}(0))^{2}}{2u^{2}_{x}(0)})$.

For the second setup where the measurement operation and the potential change operation are applied in the same direction. Now without the loss of generality we can
assume that both operations are taken along the $x$ direction. Then the $x$ direction distribution after the measurement can be obtained as,
\begin{equation}\small
\mathcal{X}(x|x_{m})=\frac{1}{\sqrt{2\pi\frac{u_{x}^{2}(0)u_{x,m}^{2}}{u_{x}^{2}(0)+u_{x,m}^{2}}}}\exp(-\frac{(x-\frac{x_{m}u_{x}^{2}(0)+d_{x}(0)
u_{x,m}^{2}}{u_{x}^{2}(0)+u_{x,m}^{2}})^{2}}{2\frac{u_{x}^{2}(0)u_{x,m}^{2}}{u_{x}^{2}(0)+u_{x,m}^{2}}}).
\end{equation}
In this situation, the distribution along the $y$ direction is $\mathcal{Y}(y,0)=\frac{1}{\sqrt{2\pi u^{2}_{y}(0)}}\exp(-\frac{(y-d_{y}(0))^{2}}{2u^{2}_{y}(0)})$.

Next we introduce the concept of  \textit{relative entropy}. Suppose we measure the particle's position along the $y$ direction, then the expression of the relative entropy is
given by  \cite{D. Abreu11},
\begin{equation}\small
I_{m}(y_{m})=\frac{1}{2}\ln(1+\frac{u_{y}^{2}(0)}{u_{y,m}^{2}})+\frac{u_{y}^{2}(0)}{2(u^{2}_{y}(0)+u_{y,m}^{2})}(\frac{(y_{m}-d_{y}(0))^{2}}
{u_{y}^{2}(0)+u_{y,m}^{2}}-1),
\end{equation}
A similar expression $I_{m}(x_{m})$ can be obtained when the measurement is applied along the $x$ direction.

Now turning to the model consisting of the magnetic field coupled Brownian particles \cite{J. B. Taylor1962, Bastida2012, R. M. Velasco2013}. We assume that at time $t=0$, the $x$ and $y$ potentials are given by $k_{x}(0)=k_{y}(0)=1$, $x_{0}=y_{0}=0$. The system initial distribution before the measurement is the product of the two thermal states in $x, y$
direction corresponding to the initial potentials ($u_{x,0}=u_{y,0}=1$, $d_{x,0}=d_{y,0}=0$). The system free energy may be calculated in a similar way \cite{D. Abreu11},
 $F=-\ln(\int\exp(-V(x)))dx=-\frac{1}{2}\ln(\frac{2\pi}{k})$,  here $\Delta F$ stands for the equilibrium free energy difference. Further,  we assume that the measurement is applied instantaneously, and the magnetic field is applied for a finite period of time.  Our purpose is to study the extracted work during this finite time duration.
 For simplicity, the final condition of the total system is set as the same as the initial condition of the system (Therefore, one has $\Delta F=0$).

 To begin with,  for the first setup, we consider the effect of the outcomes of different  measurements on the work extraction.  We calculate the work $W$ and
 relative entropy $I_{m}$ with different measurement results $y_{m}$. If the applied measurement is very accurate ($u_{y,m}=0.01$), after the measurement,
 the $y$ direction distribution can be obtained as,
\begin{equation}
\mathcal{Y}(y|y_{m})=\frac{1}{\sqrt{2\pi u_{y,m}^{2}}}\exp(-\frac{(y-y_{m})^{2}}{2u_{y,m}^{2}}).
\end{equation}

%It is easily found that after this very accurate measurement, the system distribution along the $y$ direction approaches the $\delta$-pulse near the measurement position $y_{m}$.
The value of the magnetic field is set as $L(qB/c\gamma)=0.5$.
\begin{figure}[htbp]
\begin{center}
\includegraphics[width=0.5\textwidth]{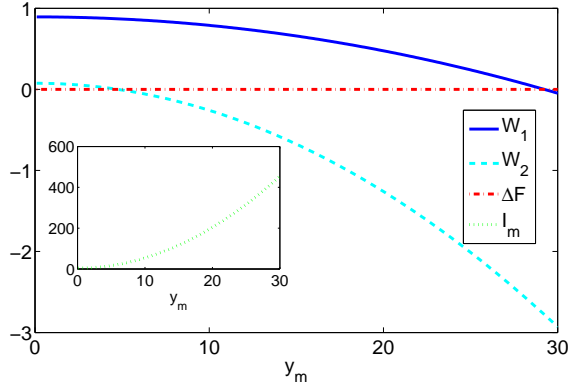}
\end{center}
\caption{\label{fig1} (Color online): $W_{1}$, $W_{2}$, $\Delta F$, and $I_{m}$ with $y_{m}$. Two different $x$-direction potentials are chosen, 1). $W_{1}$, $V_{x}(t)=\frac{1}{2}(0.5+0.5\cdot\cos(0.5t))x^{2}$, $t=4\pi$. 1). $W_{2}$, $V_{x}(t)=\frac{1}{2}(1-0.5\cdot\sin(0.5t))x^{2}$, $t=4\pi$. the potential in the $y$ direction is constant, $V_{y}=\frac{1}{2}y^{2}$. The thermal relaxation coefficients are $N=1$, the measurement is performed  in $y$-direction, parameters are $u_{y,0}=u_{x,0}=1$, $d_{y,0}=d_{x,0}=0$, and $u_{y,m}=0.01$, the magnetic field takes the value $L(qB/c\gamma)=0.5$.}
\end{figure}
Fig.~\ref{fig1} shows that, when the outcome of the position measurement is close to the origin point ($y_{m}$=0), no work can be extracted from the environment. When the measurement outcomes exceeds a certain threshold value, however, the applied work value becomes negative, which means that the work can be extracted from the
environment by employing the continuous potential change. Although $W(y_{m})$ is smaller than the free energy difference $\Delta F$, the generalized second law $W(y_{m})+I_{m}(y_{m})\geq\Delta F=0$ is still satisfied \cite{D. Abreu11, M. Bauer12, Granger11PRE}, and this process above is a thermodynamically validated process.

\begin{figure}[htbp]
\begin{center}
\includegraphics[width=0.5\textwidth]{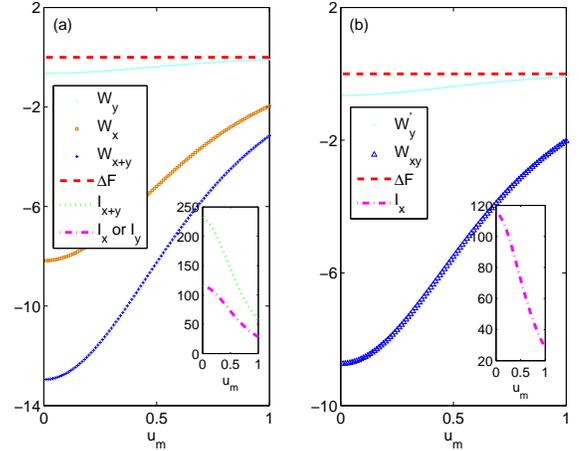}
\end{center}
\caption{\label{fig2} (Color Online): (a)  $W_{y}$, $W_{x}$, $W_{x+y}$, $\Delta F$, $I_{x}$ or $I_{y}$, and $I_{x+y}$ with measurement accuracy $u_{m}$ at time $t=4\pi$. (1). $W_{y}$, the measurement is along the $y$ direction, the position measurement result $y_{m}=15$. (2). $W_{x}$, the measurement is along the $x$ direction, the position measurement result $x_{m}=15$. (3). $W_{x+y}$, the measurements are along the $x$ and $y$ directions, the position measurement results $x_{m}=y_{m}=15$. The potential of the $x$ direction is $V_{x}(t)=\frac{1}{2}(1-0.5\cdot\sin(0.5t))x^{2}$, the constant potential on the $y$ direction, $V_{y}=\frac{1}{2}y^{2}$. (b) $W_{y}^{'}$, $W_{xy}$, $\Delta F$, $I_{x}$ with measurement accuracy $u_{m}$ at time $t=4\pi$. (1). $W_{y}^{'}$, the potential of the $x$ direction is constant, $V_{x}=\frac{1}{2}x^{2}$. (2). $W_{xy}$, the potential of the $x$ direction is $V_{x}(t)=\frac{1}{2}(1-0.5\cdot\sin(0.5t))x^{2}$. The potential on the $y$ direction,  $V_{y}(t)=\frac{1}{2}(1-0.5\cdot\sin(0.5t))y^{2}$. The measurement is along the $x$ direction, the position measurement result $x_{m}=15$. For (a) and (b), the thermal relaxation coefficients are set $N=1$, the following parameters are set as $u_{y,0}=u_{x,0}=1$, $d_{y,0}=d_{x,0}=0$, the magnetic field is $L(qB/c\gamma)=0.5$.}
\end{figure}

%%%%%%%%%%%%%%%%
For the second setup, we assume that both measurement and potential change are performed in the x direction.  As a comparison, we also discuss the process involving
a measurement in the y direction while the potential change is still in x direction.   Other parameters are set equally in these two different setups.  The work extraction for
 the both setups is shown in Fig.~\ref{fig2}.~(a). In both setups, the work extraction is always possible even the measurement and potential change directions are different.
 However, more work can be extracted when the measurement and potential change directions coincide (all in x-direction).  Specifically, the extract work from the two
spatially separated operations is realized when the measurement accuracy reaches a certain value. Interestingly, it is shown that in both cases the generalized second
law is satisfied.

%{\bf to be added} This suggests we may extract more work when two independent measurement processes are applied..........

%%%%%%%%%%%%%%%%%%%%

Moreover, as the third setup which combines the above setups, we discuss the work extraction that the measurement operations are executed along the $x$ and $y$ directions. The potential change is along the $x$ direction, from Fig.~\ref{fig2}.~(a) above, the extracted work $W_{x+y}$ is larger than the measurement operation is applied in one direction solely.

Next, we study the extracted work of the fourth setup. The first setup is included here to compare with the work value $W_{xy}$ from the fourth setup. As shown in Fig.~\ref{fig2}.~(b), more work is extracted when potential changes take place in both $x$ and $y$ directions.
%\begin{figure}[htbp]
%\begin{center}
%\includegraphics[width=0.5\textwidth]{x+y.eps}
%\end{center}
%\caption{\label{fig3} (color online): $W_{x+y}$, $\Delta F$, and $I_{m}$ with measurement accuracy $u_{m}$ at time $t=4\pi$. The measurement operations are executed along the $x$ and $y$ directions, the position measurement results $x_{m}=y_{m}=15$. The potential of the $x$ direction is $V_{x}(t)=\frac{1}{2}(1-0.5\cdot\sin(0.5t))x^{2}$, the constant potential on the $y$ direction, $V_{y}=\frac{1}{2}y^{2}$. The thermal relaxation coefficients are set, $N=1$, parameters are set, $u_{y,0}=u_{x,0}=1$, $d_{y,0}=d_{x,0}=0$, the magnetic field satisfies, $L(qB/c\gamma)=0.5$.}
%\end{figure}

\begin{figure}[htbp]
\begin{center}
\includegraphics[width=0.45\textwidth]{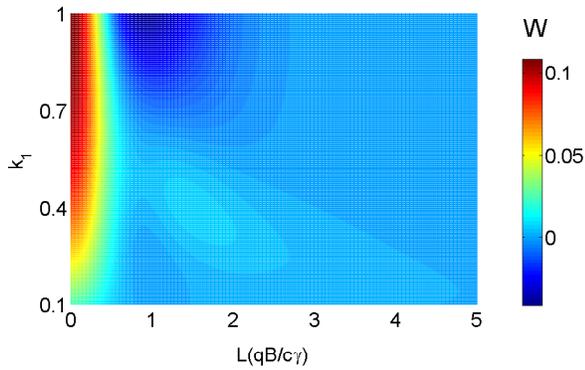}
\end{center}
\caption{\label{fig3} (Color Online): The extracted work $W$ with the change of the magnetic field and the potential change frequency. The potential of the $x$ direction is $V_{x}(t)=\frac{1}{2}(1-0.5\cdot\sin(k_{1}t))x^{2}$, the evolution time is the half integer period $k_{1}t=\pi$, the constant potential on the $y$ direction, $V_{y}=\frac{1}{2}y^{2}$. The thermal relaxation coefficients are set, $N=1$, the measurement is done on the $y$ direction, parameters are set, $u_{y,0}=u_{x,0}=1$, $d_{y,0}=d_{x,0}=0$, $u_{y,m}=0.01$, and $y_{m}=4$.}
\end{figure}

%%%%%%%%%%%%%%
In addition,  we study the effect of the magnetic field and the potential change frequency on the work extraction. Fig.~\ref{fig3} shows the extracted work during a half integer period ($k_{1}t=\pi$), and $\Delta F=0$. The potential change is taken along the $x$ direction, while the measurement is along the $y$ direction. When the magnetic field is weak, the coupling between the two direction Brownian particles motion is weak, and the motion in these two directions ($x, y$) can be approximately seen as two separated motions, so the measurement along the $y$ direction has little effect on the $x$ direction dynamics. It is seen that in this weak-coupling limit the potential change operation
cannot extract the work from the reservoir. When the magnetic field coupling strength is relatively large, the $x$ and $y$ direction dynamics are coupled, and one can extract work from the reservoir ($W<0$). However, it is worthwhile to note that, if the magnetic field becomes too strong (exceeding a threshhold), the applied work approaches zero. This is because the magnetic field interaction dominates all the interactions of the system rendering the external potentials to be ignorable.  When there is no external force applied, the work done on the system is close to zero.  As shown in Fig.~\ref{fig3}, one can find that in our continuous modulation case, the maximum work can be obtained for a certain magnetic field parameter. Of course, this value is affected by the frequency of the potential change.

\begin{figure}[htbp]
\begin{center}
\includegraphics[width=0.5\textwidth]{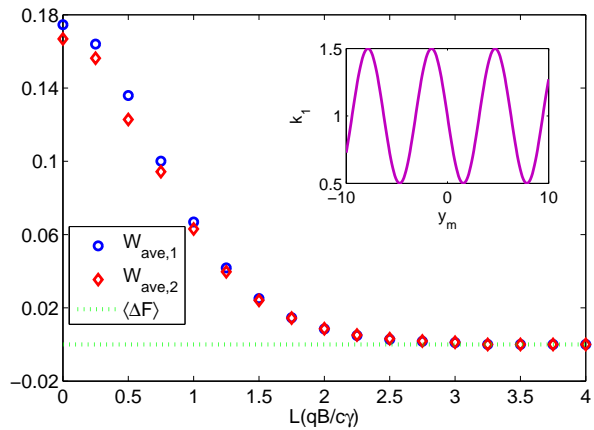}
\end{center}
\caption{\label{fig4} (Color online): $W_{ave}$, and $I_{ave}$ with $L(qB/c\gamma)$ at time $t=2\pi$. These two quantities $W_{ave}$ and $I_{ave}$ average over different $y$ directional measurement outcomes $y_{m}$. The potential of the $x$ direction is, $V_{x}(t)=\frac{1}{2}(1-0.5\cdot\sin(k_{1}\cdot t)) x^{2}$, and $W_{ave,1}$, $k_{1}=1$; $W_{ave,2}$, $k_{1}=1-0.5 y_{m}$, seen in the inset. The constant potential on the $y$ direction, $V_{y}=\frac{1}{2}y^{2}$. The thermal relaxation coefficients are set, $N=1$, the measurement is done on the $y$ direction, parameters are set, $u_{y,0}=u_{x,0}=1$, $u_{y,m}=0.1$, $d_{y,0}=d_{x,0}=0$.}
\end{figure}

Now we will consider the average over the outcomes of the position measurement. The mean total work $W_{\rm ave}=\int p(y_{m})W(y_{m})d y_{m}$, the mutual information $I_{ave}=\int p(y_{m})I_{m}(y_{m})d y_{m}$ where  $p(y_{m})$ is the probability of the position measurement with outcome $y_{m}$, it is interesting to consider the feedback
effect on the work extraction. The potential is chosen as $V_{x}(t)=\frac{1}{2}(1-0.5\cdot\sin(k_{1}\cdot t))x^{2}$, for $W_{ave,1}$, $k_{1}=1$, and for $W_{ave,2}$, the parameter $k_{1}$ is related to the measurement outcome $y_{m}$ by, $k_{1}=1-0.5\sin(y_{m})$, the time duration is $k_{1}\cdot t=2\pi$, so the system free energy difference $\Delta F$ is zero, see Fig.~\ref{fig4}. The average work value with the different magnetic fields  is shown, with $W_{ave,(1,2)}\geq\langle\Delta F\rangle$, we find that when the potential stiffness is changed with the measurement outcome, the applied work value $W_{ave,2}$ is smaller than that of not related to the measurement outcome $W_{ave,1}$, so the feedback is beneficial to the extracted work. When the magnetic field dominates the interaction, the work done by the external part approaches zero. We also calculate the mutual information $I_{ave}=$2.3076 from the integral above in our chosen protocol, which is the exact value of $I_{ave}=\frac{1}{2}\ln(1+\frac{u_{y,0}^{2}}{u_{y,m}^{2}})$ \cite{D. Abreu11}.

\section{conclusion}\label{IV}
In summary, we have studied the work extraction in a model consisting of the magnetic field coupled to Brownian particles. We find that, by employing the information contribution, we can extract work from the reservoir even in the case where the measurement operation and the potential change operation are spatially separated. In particular, we show that more work can be extracted when two independent measurements are considered. In all the processes considered in this paper the generalized second law is obeyed. Furthermore, the extracted work for the different magnetic fields and the potential change frequencies as well as the measurement accuracies are discussed. In addition, the feedback effect on the work extraction is discussed, the applied work in the feedback case is shown to less than that in the no-feedback case. Physically, it implies that  the feedback process can be used to gain more information from the system. It will be interesting to consider the work extraction in a case where more general measurement
processes are considered. This will be the topic of our future investigation.

Note Added: After our manuscript was in print, we became aware of a paper available in arXiv (arXiv:1405.1461v1), which also considers a similar topic as ours.

\section*{Acknowledgement}
This work is supported in part by the 10000-Plan of Shandong province, and the National High-Tech Program of China Grants No. 2011AA010800 and 2011AA010803, NSFC Grants No. 11174177 and 60725416 (T.C. and X. B. W.). T.C. is grateful to the financial support from China Scholarship Council No. 201206210176. T.Y. acknowledges Grant support from the NSF PHY-0925174 and DOD/AF/AFOSR No. FA9550-12-1-0001.
{}

\end{document}